\documentclass[conference]{IEEEtran}
\usepackage{lipsum,hyperref,booktabs,braket}
\usepackage{algorithm}
\usepackage[noend]{algpseudocode}

\IEEEoverridecommandlockouts
\usepackage{cite}
\usepackage{amsmath,amssymb,amsfonts}
\usepackage{graphicx}
\usepackage{textcomp}
\usepackage{xcolor}

\newtheorem{definition}{Definition}

\begin{document}

\title{An Earth Mover's Distance Based Graph Distance Metric For Financial Statements 
\thanks{For reproducibility, the source code and synthetic data are publicly available at \url{https://github.com/snoels/earth-movers-graph-distance-metric}.}
}

\author{
\IEEEauthorblockN{Sander Noels\IEEEauthorrefmark{1}\IEEEauthorrefmark{2} Benjamin Vandermarliere\IEEEauthorrefmark{2} Ken Bastiaensen\IEEEauthorrefmark{2} Tijl De Bie\IEEEauthorrefmark{1}}
\IEEEauthorblockA{\IEEEauthorrefmark{1}Department of Electronics and Information Systems, Ghent University
\\Ghent, Belgium
\\\{sander.noels, tijl.debie\}@ugent.be}
\IEEEauthorblockA{\IEEEauthorrefmark{2}Silverfin
\\Ghent, Belgium
\\\{sander.noels, benjamin.vandermarliere, ken.bastiaensen\}@silverfin.com}
}

\maketitle

\begin{abstract}
Quantifying the similarity between a group of companies has proven to be useful for several purposes, including company benchmarking, fraud detection, and searching for investment opportunities. This exercise can be done using a variety of data sources, such as company activity data and financial data. However, ledger account data is widely available and is standardized to a large extent. Such ledger accounts within a financial statement can be represented by means of a tree, i.e. a special type of graph, representing both the values of the ledger accounts and the relationships between them. Given their broad availability and rich information content, financial statements form a prime data source based on which company similarities or distances could be computed.

In this paper, we present a graph distance metric that enables one to compute the similarity between the financial statements of two companies. We conduct a comprehensive experimental study using real-world financial data to demonstrate the usefulness of our proposed distance metric. The experimental results show promising results on a number of use cases. This method may be useful for investors looking for investment opportunities, government officials attempting to identify fraudulent companies, and accountants looking to benchmark a group of companies based on their financial statements.
\end{abstract}


\section{Introduction}\label{section1}

A financial statement provides a concise and comprehensive overview of a company's financial position and acts as a good predictor of future performance \cite{nagy1994factors}. This encourages investors and government regulators to compare financial statements when making investment decisions or detecting fraud \cite{yang2013balance}. Aside from looking for unusual companies, company benchmarking can provide a comprehensive overview of the industry. However, the manual comparison and analysis of financial statements is a tedious and time consuming task. This creates the need for a data-driven and automated solution that reduces the processing time \cite{cong2014impact}.

In the past, several attempts have been made to define company similarity. This appears to be beneficial for company classification and fraud detection purposes \cite{jan2018effective,yang2013balance,kanapickiene2015model,yang2019companyclassification}. However, the previously proposed methodologies only analyze a portion of the information present in a financial statement: either the values on the ledger accounts or the structural relationship between the ledger accounts present within a financial statement. This inspires the idea of considering both the structural properties as well as the value information to quantify the similarity between two companies.

In this paper we propose a new graph distance metric based on the earth mover's distance (EMD) \cite{rubner2000earth}. The metric allows one to quantify the similarity between two financial statements, taking into account both structure and value information. We demonstrate the effectiveness of this distance metric compared to the methodologies proposed in earlier studies. In this paper we concentrate on the balance sheet component of a financial statement. This work can be extended to other components of the financial statement and is by no means exhaustive for financial applications alone. This distance metric introduces a data-driven way of computing company similarities, which could be beneficial for investors looking for investment opportunities, government officials attempting to identify fraudulent companies, and accountants looking to benchmark a group of companies.

Our main contributions are summarized as follows:
\begin{itemize}
  \item We propose a new graph distance metric that takes into account both structure and value information that allows one to compute the similarity between two financial statements.
  \item We provide a detailed description of how a graph distance metric can be applied to financial statements.
  \item We demonstrate how the distance metric can be used for dimensionality reduction purposes and apply it to t-SNE.
  \item We conduct a comprehensive experimental study using real-world financial data to demonstrate the usefulness of our proposed distance metric when computing the distance between the financial statements of two companies.
 \end{itemize}

The remainder of the paper is structured as follows. Section \ref{section2}, gives a summary of the previous related work. Section \ref{section3} introduces the graph distance metric for financial statements. Section \ref{section4} discusses how to determine the weight function required by our distance metric. In section \ref{section5}, we provide an experimental evaluation of our proposed method, and section \ref{section6} concludes this work and gives an overview of possible future studies.

\section{Related Work}\label{section2}

Financial statements comprehensively portray the operating activities and financial performance of a company. Typically financial statements include balance sheets, statements of profit or loss, and reconciliations. Because financial statements provide a succinct and all-encompassing summary of a company's financial situation, investors consider it as a good indicator of company performance \cite{nagy1994factors}. Hopkins \cite{hopkins1996effect} claims that the stock price judgment of financial analysts is influenced by the assessment of the balance sheet. This means that if well-performing companies are known, companies with similar balance sheets should perform equally well. 

Aside from the similarity of financial statements, dissimilar financial statements can also provide useful information.
One instance where this is demonstrated is fraud detection \cite{jan2018effective,kanapickiene2015model}. Companies sometimes manipulate financial figures to gain access to long-term debt financing or to boost stock prices. A company distance metric that enables supervisory bodies to detect these unconventional financial statements is thus extremely valuable. Additionally, deviations may also reveal unique company characteristics. These distinct characteristics could imply that a company is uniquely positioned, which could indicate a strong investment opportunity \cite{yang2013balance}. 

Furthermore, several studies \cite{yang2013balance,de2011benefits} have suggested that organizations with similar business activities should have similar financial statements. This idea is confirmed by Yang et al. \cite{yang2019companyclassification}, where they provide evidence that a company distance metric can effectively identify industry boundaries. 

With the advancement of information technology, there is an increased interest in utilizing technology to improve the information processing speed \cite{cong2014impact}. This emphasizes the need for a company similarity metric that can quantify company similarity in a data-driven fashion. 

Several attempts have been made to identify similar companies. Industry classification standards allow companies to be classified into homogeneous categories with the assumption that companies within the same group display similar characteristics. The statistical classification of economic activities in the European community (NACE), is the classification standard of the European Union. This classification standard can be compared with the Standard Industry Classification (SIC) and North American Industry Classification System (NAICS). It is well-recognized that industry classification aids in company analysis when compared to simply considering firm size \cite{kahle1996impact}. 

However, industry classification schemes have their limitations. One of the drawbacks is that classification systems do not evolve at the same rate as the market conditions, making it difficult to classify new industries \cite{fan2000relatedness}. Another drawback is the lack of uniform classification standards, which results in a different company classification depending on the industry classification standard being used \cite{yang2019companyclassification}. This implies that we cannot merely distinguish organizations based on their size and industry classification. Companies might differ greatly even within the same industry, necessitating the development of a cross-industry similarity metric.

Several attempts have been made to determine the similarity of companies based on their financial statements. Financial ratios are one way to figure out how similar two companies are \cite{kanapickiene2015model}. Financial ratios rely on extracting data from financial statements in order to derive meaningful numerical values that reflect the current operating activities or financial performance of a company. This means that in order to compare the financial performance of companies, a set of financial ratios should be selected. As a result, selection bias can enter the process. Another restriction is that companies may attempt to apply window dressing to improve their financial ratios. Financial analysts must be wary of these practices that artificially inflate the solvency or liquidity of a company.

Another line of research tries to tackle this problem by looking at the financial statements as a whole. Brown, Ma, and Tucker \cite{brown2021financial} represent a company as a vector where each element represents a ledger account value. They define the similarity between two companies as the cosine or Mahalanobis distance between these vectors. This yields a numerical value that expresses how similar two companies are. This strategy, however, does not take into account the structure of a financial statement. More specifically, the relatedness and hierarchical position of ledger accounts within a financial statement have no effect on the distance measure. This means that two ledger accounts that are closely related, e.g. \emph{land} and \emph{buildings}, have the same effect on the distance metric as two ledger accounts that are completely unrelated.

The paper of Yang and Cogill \cite{yang2013balance}, which acts as foundation for our research, advocates for using the structural properties of the ledger accounts present in a balance sheet. They developed a tree edit distance-based algorithm that considers companies to be similar if their balance sheet structures are similar. Regrettably, this strategy only evaluates the structure of a companies' balance sheet. This means that the account values on the ledger accounts are not taken into consideration.

Understanding the relationship between assets, liabilities, expense, and revenue structure is crucial to understanding the financial situation of a company \cite{yang2013balance}. This inspires the idea of considering both the balance sheet structure as well as the values on the ledger accounts when determining the similarity between two companies.

\section{Tree Distance Metric}\label{section3}

This section starts with introducing the tree representation of a financial statement, followed by the motivation for our tree distance metric. Subsequently, a visual representation of our method is provided, accompanied by the mathematical description of our method.

\subsection{Financial Statements as a Graph}\label{FSG}
As stated by Yang and Cogill \cite{yang2013balance}, a vertex-labeled tree is a natural representation of the ledger accounts present within a financial statement. 

\begin{figure}[htbp]
\centerline{\includegraphics[width=0.5\textwidth]{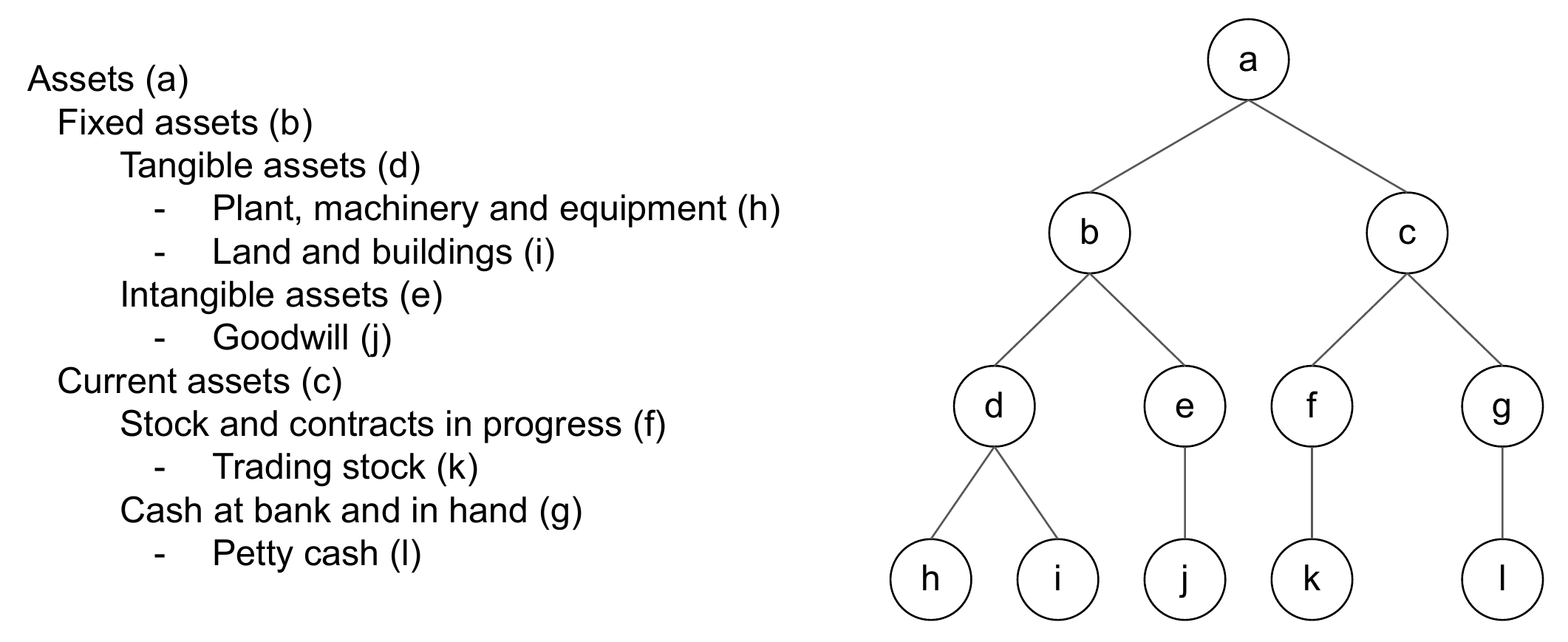}}
\caption{Left: Assets subsection of the balance sheet. Right: A vertex-labeled tree representation of the assets subsection of the balance sheet.}
\label{fig1}
\end{figure}

As example, we consider the \emph{assets} section of a balance sheet. The \emph{assets} section can be divided into \emph{fixed} and \emph{current assets}. Consequently, a ledger account can be subdivided into more detailed accounts. As shown in figure \ref{fig1}, the ledger account \emph{plant, machinery and equipment} falls under the \emph{fixed assets} section, which can be further subdivided into \emph{tangible} and \emph{intangible assets}. A ledger account can also be a part of the \emph{current assets} section, which is subdivided into \emph{stocks and contracts in progress} and \emph{cash at bank and in hand}. It is worth noting that the vertex-labeled representation of a balance sheet is not limited to this specific example. A subset of ledger accounts and their reciprocal relation are given for exemplary purposes. 

This representation method clearly preserves the structural property of a financial statement. Besides balance sheets, statements of profit and loss can also be represented by this structural fashion. This means that a financial statement of a company can be represented by a vertex-labeled tree where the vertex labels are the ledger account names. The tree of all possible financial accounts hierarchically structured within a financial statement serves as the general structure of a financial statement, allowing us to represent every company. 

In this paper, we use a vertex-weighted tree to represent a company. This means that each node in the tree is given a weight. This weight is assigned to a specific node based on its ledger account value; more specifically, it equals the node's relative importance depending on its ledger account value. We refer to the function that assigns a weight to a node as the \emph{weight function} $w$ as described in section \ref{section4}.

\subsection{Motivation}
Understanding the interaction between assets, liabilities, expense, and revenue structure is directly related to understanding a company's financial position \cite{yang2013balance}. This motivates us to develop a company distance metric that takes into account the structure of the ledger accounts used by a company's financial statement, as well as the relative distribution of the ledger account values. Our motivation is based on the assumption that companies are similar if they have a similar balance sheet structure, as well as a similar weight distribution over their balance sheet.

\begin{figure*}[t]
    \centering
    \includegraphics[width=\textwidth]{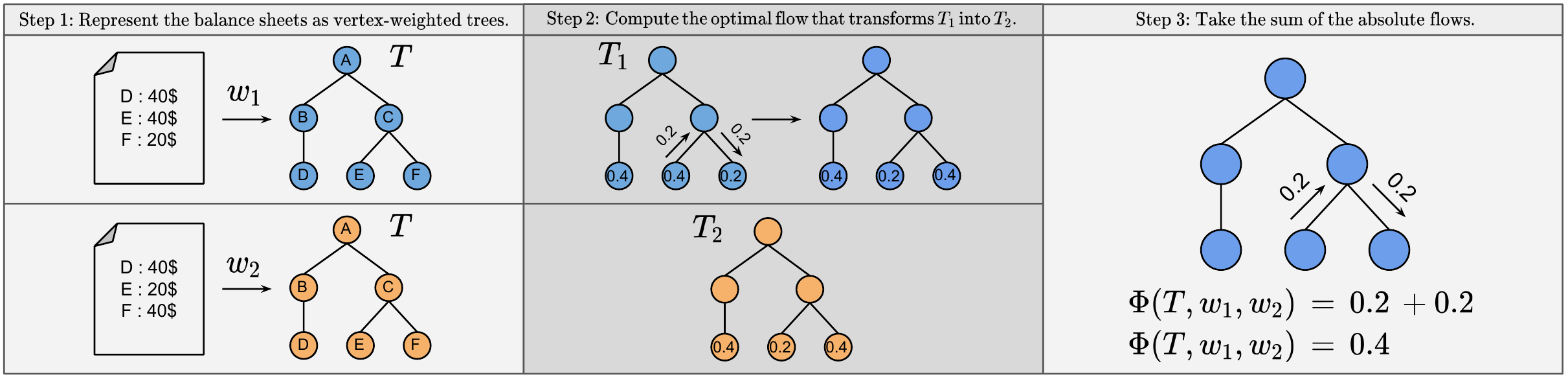}
    \caption{Graphical representation of how our proposed distance metric calculates the distance between two companies. Step 1 shows how the company specific weight function $w$ transforms the general tree structure $T$ into a company specific vertex-weighted tree. Step 2 computes the optimal edge-flows so that $T_1$ and $T_2$ become identical. Step 3 takes the absolute sum of the optimal edge-flows which represents the distance between two companies.}
\label{fig2}
\end{figure*}

First of all it is important that our metric understands the reciprocal relatedness of two ledger accounts. Two ledger accounts located under the \emph{land and buildings} node should be considered as more related, whilst two other ledger accounts - that are not located under the same parental node - should be considered as less related. It is here where the role of structure information comes into play. This goes beyond the approach of Brown, Ma, and Tucker \cite{brown2021financial}, where they neglect the general hierarchy of the balance sheet, and only take into account the values of the ledger accounts. 

Although two companies might be similar structure-wise, they should not be evaluated as similar if their ledger account value distribution is completely different. Consider the situation where there are two companies with very similar balance sheet structures. For one company, the highest ledger account weight could be located on the \emph{buildings} node, while another company might not own any property. Despite having similar balance sheet structures, these companies should not be considered as very similar. This shows that company similarity should also be influenced by the ledger account values located within a financial statement.

\subsection{Tree Distance Metric}

Every company can be represented by the same generic tree (see \ref{FSG}). Let us denote this generic tree as $T = (V,E)$, where $V$ is a set of $|V| = n$ nodes, and $E$ is the set of edges. Subsequently, we define the company specific weight function $w \colon V \mapsto \mathbb{R}$ that transforms the generic tree into a company specific tree by assigning a weight to every node. The generic tree $T$ and the company specific weight function $w$ allow us to map every company balance sheet to a company specific tree representation.

Let $T_{1} = (V,E,w_{1})$ be the tree representation of company one and $T_{2} = (V,E,w_{2})$ the tree representation of company two. Based on our motivation, we want to quantify the similarity between two companies related to the structure and value information of their ledger accounts. We define the similarity between two vertex-weighted trees as the total cost of shifting weights over the edges of $T_1$ in order to become identical to $T_2$. A company that is slightly different from another company based on their balance sheet structure and ledger account value distribution, does not require a lot of weight shifts. On the other hand, very dissimilar companies require a lot of weight shifts.

This distance metric is based on the EMD \cite{rubner2000earth} and calculates the minimal amount of weight shifts over the edges of $T_{1}$ in order to become identical to $T_{2}$. In this example, tree 1 acts as the source tree, while tree 2 acts as the sink tree. It is worth noting that this distance is symmetric. Figure \ref{fig2} shows a graphical example of how the distance metric works. \\
\\
\noindent This brings us to the formal definition of our graph distance metric:

\begin{definition}[Earth Mover's Distance Based Graph Distance Metric]
Given an undirected graph $T = (V,E)$ with $|V| = n$ and two weight functions $w_{1} \colon V \mapsto \mathbb{R}$ and $w_{2} \colon V \mapsto \mathbb{R}$ where $\sum_{i=1}^{n}w_{1}(v_{i}) = \sum_{i=1}^{n}w_{2}(v_{i})$. Consider $T_{1} = (V,E,w_{1})$ as the source graph where $p_{i} = w_1(v_{i})$ is the production weight associated with node $i$, also consider $T_{2} = (V,E,w_{2})$ as the sink graph where $c_{i} = w_2(v_{i})$ is the consumption weight associated with node $i$. Then the distance between the graphs $T_1$ and $T_2$, denoted as $\phi(T,w_{1},w_{2})$, is defined as the minimum amount of total weight allocation that has to be shifted over the edges of $T_1$ in order to become identical to $T_2$.
\end{definition} 
\hspace{0.4\textwidth}

\noindent Computing this distance can be done by
solving a linear programming problem for finding the edge flows $f_{i \rightarrow j}$ that minimize the overall cost:

$$\text{minimize } C = \sum_{(i,j)\in E} |f_{i \rightarrow j}|$$

subject to the constraint that for every node $i$:

$$\sum_{j:(i,j)\in E} f_{i\rightarrow j}=p_{i}-c_{i}.$$

This distance metric searches for the optimal flow matrix $\mathbf{F} \in \mathbb{R}^{n\times n}$ where the graph distance is defined as $\sum|\mathbf{F}|$, with the only constraint that the total flow for a node $i$ is equal to the production $p_{i}$ minus the consumption $c_{i}$. Since the absolute value of a flow is a non-linear function, objective $C$ can be transformed into a linear function \cite{boyd2004convex} by introducing the new variable $g_{ij}$:

$$\text{minimize } C = \sum_{(i,j)\in E} g_{ij}$$

subject to the constraints:

$$g_{ij} \geq f_{i \rightarrow j},$$
$$g_{ij} \geq - f_{i \rightarrow j}.$$

In the following section we elaborate on the determination of the weight function $w$ and the general tree representation $T$.

\section{Determining the weight function}\label{section4}

In this paper we subdivide the tree representation of a balance sheet into a set of sub-trees. More specifically we divide a balance sheet into four different trees: the debit active tree, the credit active tree, the credit passive tree, and the debit passive tree. This technique eliminates the possibility of transferring weights from the active to the passive side and vice-versa. Transferring weights between the debit and credit sides of the active or passive tree is also prohibited. Allowing this goes against basic principles of accounting.

The weight function we propose in this paper is $w(v_{i}) = \frac{b_{i}}{\sum_{i=1}^{n}b_{i}}$ where $b_i$ represents the ledger account value of node $i$. The weight assigned to a node through this weight function equals to the relative importance of a node in the sub-tree. As a result, the node weights are easily explainable. The node weights can be explained as follows: node weights where $w_1(v_i) > w_2(v_i)$ represent the situation where the ledger account $i$ is more important for company one compared to company two, node weights where $w_1(v_i) < w_2(v_i)$ represent the opposite, and node weights where $w_1(v_i) = w_2(v_i)$ represent the situation where both companies consider the ledger account $i$ as equally important.

When considering one general tree structure, the application of this weight function is not effective. This because of the presence of negative ledger account values. A negative ledger account value on the active side of a balance sheet indicates a credit account (e.g., a depreciation), whereas a positive value on the passive side indicates a debit account. When incorporating both debit and credit ledger account values within the active or passive tree, the vertex weights are able to expand, which results in non-explainable node weights. The assumptions mentioned above result in a logical, well-explainable weight function, in which a node weight expresses the relative importance of a ledger account within a sub-tree of ledger accounts.

Another benefit of the weight function $w$ is that it is adaptable. A company's feature vector $\mathbf{b}$, which is a vector of all booked values in the balance sheet, can be easily replaced by another feature vector. Instead of using the actually booked values, another option is to use the number of transactions associated with a certain ledger account. This allows users of this distance metric to create their own version of the metric that is tailored to their specific needs.

\section{Experiments}\label{section5}

To verify the effectiveness and applicability of the proposed graph distance metric, we conduct various experiments on real-world financial data, which allows us to interpret the properties of the graph distance metric. We conduct two different experiments where our graph distance metric is compared against several benchmark methods. Each experiment is preceded by a description of the experimental setting, followed by an evaluation of the experiment.

By conducting these experiments we want to answer 2 questions:
\begin{itemize}
  \item Does our proposed method, which considers both structure and value information, provide more information than methods that only consider one of the two?
  \item Does our graph distance metric allow one to find similar companies as well as company outliers based on their financial statements?
\end{itemize}

First we discuss the dataset, followed by the proposition of several baselines. Subsequently, we introduce two experiments where we evaluate the usefulness of our method against the baselines.

\subsection{Dataset}

We used proprietary Silverfin\footnote{\url{https://www.silverfin.com}} data to conduct the experiments. Silverfin is a Belgian scale-up focused on building an accountancy cloud service. The confidential dataset used in this paper contains the financial statement data of 1000 Belgian companies, that ended their financial year in 2019. In addition, we also have information about the commercial activities of the companies, such as the NACE codes. We constructed a set of vertex-weighted trees for every company consisting of their active and passive tree representation. The fact that this is real-word data may allow accountants using Silverfin's service to draw valuable insights. We refer to this dataset as \textbf{SILVERFIN}.

\subsection{Methods}\label{methods}

This section introduces the baselines. Since we are unaware of other methods that take into account both structure and value information of the balance sheet, we compare our proposed method against two methods that take into account one of the two. 

Consider the following methods: 

\textbf{Yang Graph Distance Metric (Y-GDM)}: The method proposed by Yang and Cogill \cite{yang2013balance} proves to be effective to detect structural changes between balance sheets. In their paper they translate the underlying company graphs into property strings and use the Levenshtein distance \cite{levenshtein1966binary} to compute the similarity between these property strings.

\textbf{Structureless Balance Sheet Distance (SBSD)}: This method represents a company as a vector where each element represents the relative importance of a certain ledger account. The distance between two vectors is computed by taking the sum of their vector subtraction. The node weights assigned to the company tree representation overlap with the vector representation of a company. 

\textbf{Random Method}: This method selects a set of random companies and ranks them in a random way. This random ranking introduces a degree of similarity based on the ranking of the companies.

\textbf{Earth Mover's Distance Based Graph Distance Metric (EMD-GDM)}: The graph distance metric proposed in this paper.

The above-mentioned methods are compared against each other in the following subsections.

\subsection{Experiment 1: Nearest Neighbors}\label{exp1}

In this part, we quantify the predictive power of the suggested distance metric. We compare our distance metric with the methods proposed in subsection \ref{methods}. More specifically, we are interested in verifying whether another balance sheet distance metric is able to increase the overlap in NACE codes between the nearest neighbors of different companies.

Algorithm \ref{exp1_alg} describes the experimental design for experiment 1. The proposed algorithm computes the average Jaccard similarity between the set of NACE codes of a company and their $k$ nearest neighbors and this for each of the metrics evaluated. This computation is done for a set of $S$ companies, after which the average is computed over all the companies in $S$. The average Jaccard similarity represents how well the NACE codes of a company and their nearest neighbors overlap.

\begin{algorithm}
\caption{Nearest Neighbors}\label{exp1_alg}
\hspace*{\algorithmicindent} \textbf{Input:} $S$(company set),$k$(number of neighbors),\\
\hspace*{8mm} $\mathbf{D}$(distance matrix) \\
\hspace*{\algorithmicindent} \textbf{Output:} $average\_jaccard\_score$
\begin{algorithmic}[1]
\Function{ComputeJaccardScore}{$S,k,\mathbf{D}$}
\State $jaccard\_score \gets 0$
\For{$s \in S$}
\State $predictions \gets GetNearestNeighbors(s,k,\mathbf{D})$
\State $z\gets 0$
\For{$p \in predictions$}
\State $s_{nace} \gets GetNaceSet(s)$
\State $p_{nace} \gets GetNaceSet(p)$
\State $jaccard \gets Jaccard(s_{nace},p_{nace})$
\State $z \gets z + jaccard$
\EndFor
\State $jaccard\_score \gets jaccard\_score + z/k$
\EndFor
\Return $jaccard\_score/Size(S)$
\EndFunction
\end{algorithmic}
\end{algorithm}

\begin{figure}[htbp]
\centerline{\includegraphics[width=0.49\textwidth]{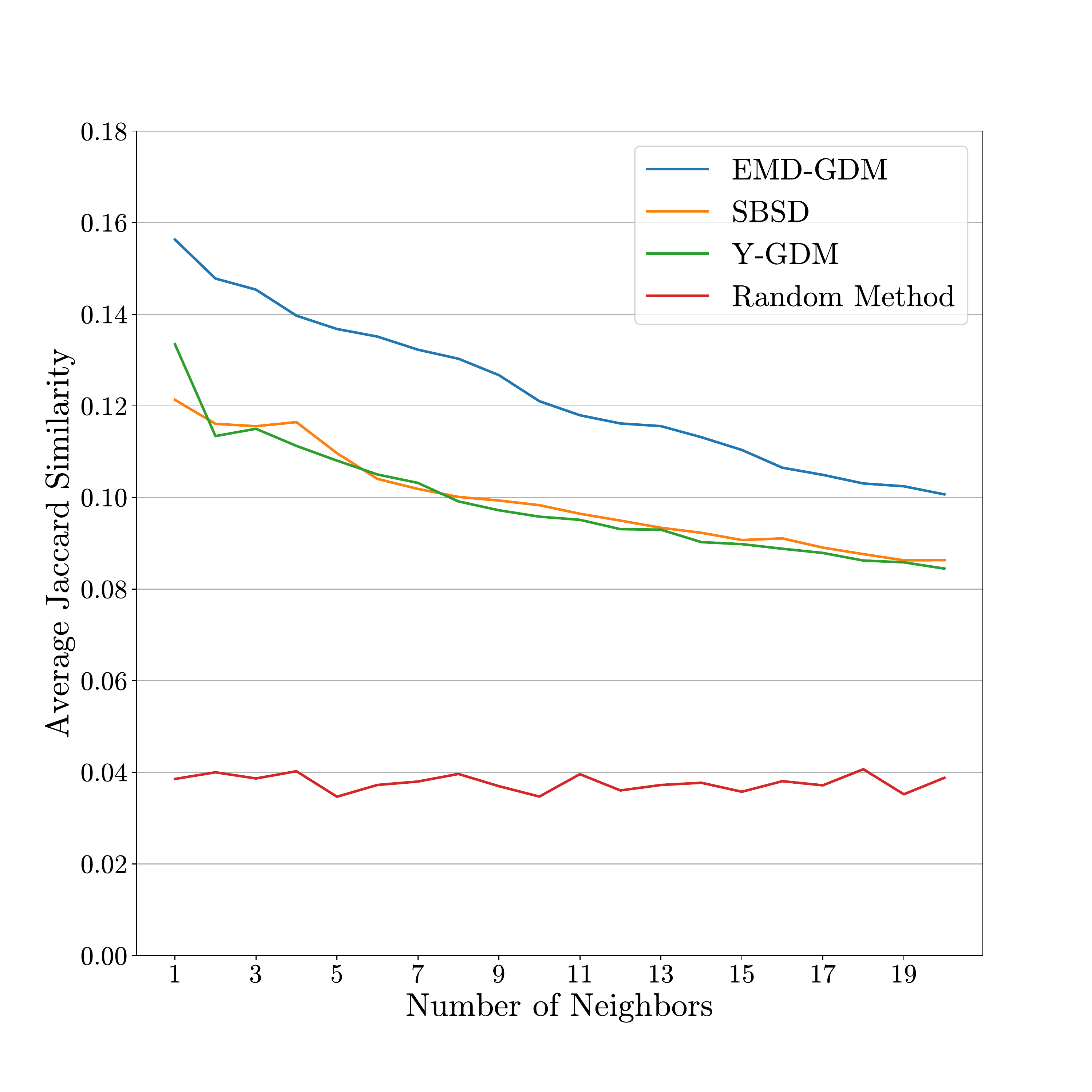}}
\caption{Average industry code Jaccard similarity between companies and their nearest neighbors.}
\label{fig3}
\end{figure}
Not all companies qualify to be part of this set $S$, because there is a large number of companies within the \textbf{SILVERFIN} dataset that do not have company neighbors that perform the same activities. Subsequently, we introduce 2 parameters that verify if a company is suitable for the experiment. Parameter $q$ specifies the number of companies that have at least one NACE code mutual with the company being verified. Parameter $r$ specifies the minimum Jaccard similarity of all the companies that have at least one mutual NACE code. In this experiment, we set $q = 20$ and $r = 0.2$. This results in approximately 400 suitable companies.

Figure \ref{fig3} depicts the performance of the different methods. The $x$-axis represents the number of chosen neighbors $k \in \braket{1,2,\dots,20}$ for whom we conducted the experiment. The average Jaccard similarity is represented on the $y$-axis. We can observe a comparable performance between the distance metric that simply considers the value distribution (SBSD) and the distance metric that simply considers the structure information (Y-GDM). Both methods highly outperform the random method that selects neighbouring companies without taking into account balance sheet information. When the random method selects a nearest neighbor there is on average a $0.04$ percent Jaccard similarity between the NACE code sets. All methods are outperformed by the novel distance metric we present in this paper. When it comes to selecting the first neighbor, the new metric performs nearly four times better as the random method. SBSD and Y-GDM perform nearly three times better than the random method. In addition, we see a decreasing trend when the number of neighbors $k$ increases. Despite this, the line of the EMD-GDM remains above the other techniques.

Selecting neighboring companies with NACE code overlap is a difficult task, given there are over 800 different ways of classifying a company. Additionally, most companies have numerous NACE codes. As mentioned in the motivational section, we expect that companies could be very similar structure-wise, but very dissimilar based on their balance sheet value distribution. This means that we expect a higher NACE code overlap between the nearest neighbors of a company when taking into account both structure and value information. We argue that this is the case because our proposed method outperforms the other two methods that only consider a part of the balance sheet information. This also implies that incorporating both structure and value information improves the distance metric's utility. The confluence of these two forms of information appears to be the key to the success of our distance metric.

\subsection{Experiment 2: Company Embedding and Local Outlier Factor}\label{exp2}

In this part, we qualitatively asses the usefulness of our proposed metric. Because we have pairwise company distances, we can translate this into a two-dimensional representation that allows us to visualize the company space. We are interested in verifying whether subgroups of companies exist in this two-dimensional space. We hypothesize that our method is able to further distinguish companies because both structure and value information is taken into account.

Additionally, we visualize a set of companies with a specific NACE code in this two-dimensional space. We analyze the companies that are located closely to each other and compare them with companies that are considered different, despite having the same NACE code. We were interested in verifying whether companies in close proximity to each other would have similar properties, as well as great distinctions between companies that were located far apart. We also hypothesize that outlier detection methods are able to detect companies within this industry that are distinctive based on their structure and value information.

For this experiment, not all companies within the \textbf{SILVERFIN} dataset are used. We exclude the companies that contain the industry NACE code ‘$70.220$’ which stands for \emph{business management consultancy}. A large number of companies have this NACE code in their industry code set, but perform very different activities. This results in almost 900 appropriate companies.

For this experiment, we use t-SNE \cite{van2008tsne} as high-dimensional data visualisation tool. Instead of using the Euclidean distance $\lVert\mathbf{x}_{i}-\mathbf{x}_{j}\lVert^{2}$ between two high-dimensional data points, we use the squared pairwise distances computed by our distance metric (see equation \ref{eqn1}). 

\begin{equation}
\label{eqn1}
p_{i|j} = \frac{exp(-d_{ij}^{2}/2\sigma_{i}^{2})}{\sum_{k\neq i}exp(-d_{ik}^{2}/2\sigma_{i}^{2})}
\end{equation}

More specifically, t-SNE makes sure that similar companies are portrayed by nearby points and dissimilar companies are portrayed by distant points with high probability. We hypothesize that companies of a similar industry locate closely together in this two-dimensional company space. The companies that are not modeled close-by should be dissimilar based on their structure or ledger account distribution. T-SNE contains a few hyperparameters that can be tuned, for this experiment we set the perplexity to $20$ and allowed the algorithm to run for $1000$ iterations.

Furthermore, we also implemented a local outlier factor (LOF) anomaly detection algorithm \cite{breunig2000lof}, which measures the local deviation of a given data point with respect to its neighbours. More specifically, this method defines the density of a certain company based on its neighbors and compares this density to the density of its neighbors. Companies that have a significantly lower density compared to their neighbors are considered outliers. The LOF-algorithm allows one to change the number of neighbors that influence the density calculation for a point, this parameter is set to $5$.

\begin{figure}[htbp]
\centerline{\includegraphics[width=0.5\textwidth]{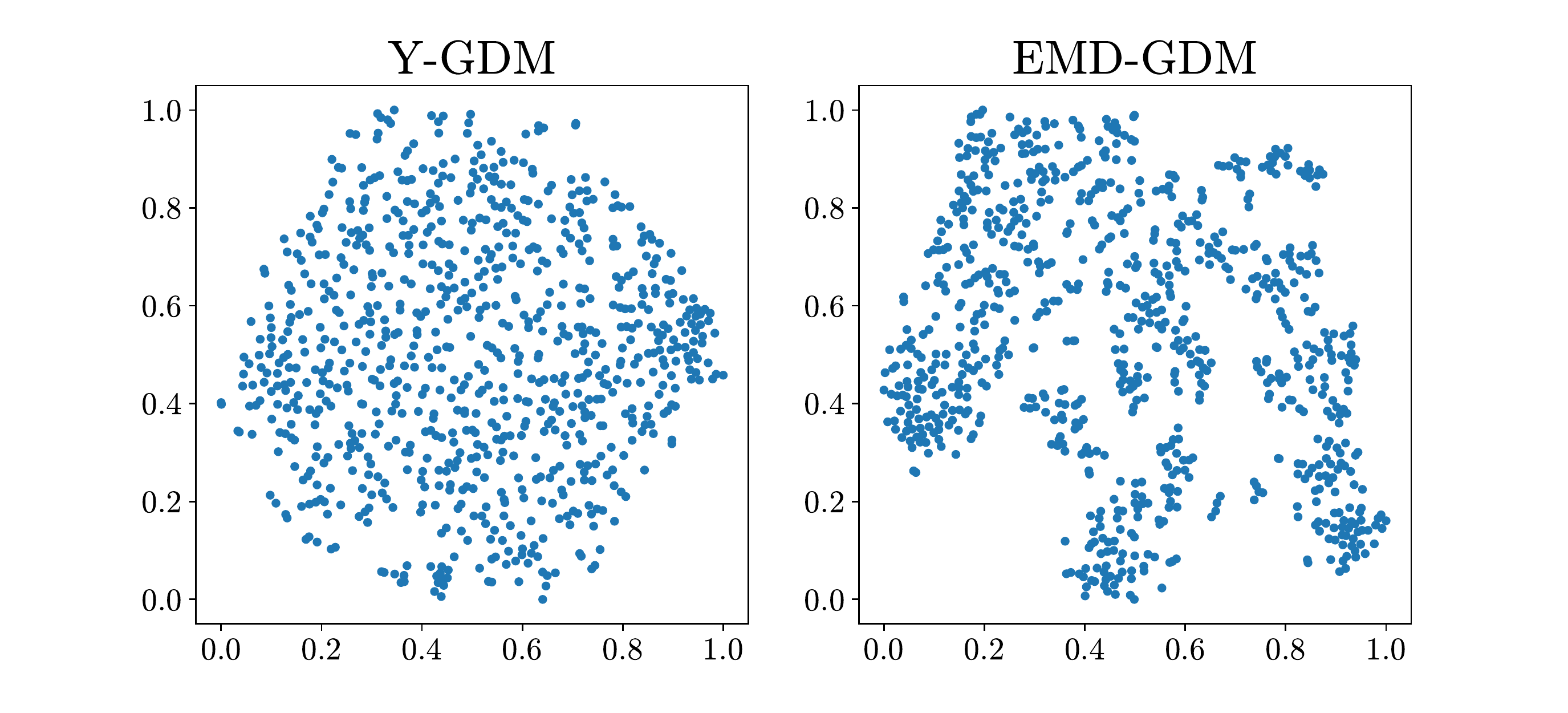}}
\caption{The t-SNE visualisation of the left graph is based on the Y-GDM. The t-SNE visualisation of the right graph is based on the EMD-GDM.}
\label{fig-tsne1}
\end{figure}

Figure \ref{fig-tsne1} shows the two-dimensional company visualisation generated by the t-SNE algorithm. Instead of using the Euclidean distances as distance metric between two high-dimensional data points, the Y-GDM is used for the left plot while the EMD-GDM is used for the right plot. T-SNE tries to capture the existing structure within the data, which means it is especially helpful for early visualization aimed at determining the degree of data separation \cite{van2008tsne}. This visualization allows us to subjectively compare the two distance metrics.

In comparison to the left plot, the right plot visually demonstrates a higher level of company separability. The left plot shows a two-dimensional company representation where most companies seem to be located equidistant. While the Y-GDM approach is almost unable to depict discrete company groups, different groups of companies emerge when our proposed distance metric is used. The perplexity values $5$, $20$, $50$, $100$, and $150$ were also evaluated. None of these perplexity values result in a clearly structured visualization for the Y-GDM distance metric, which allows us to conclude that this distance metric lacks structure. Despite the difficulty of objectively quantifying the performance of the proposed visualisation method, our suggested metric reveals clear separability of company data.

\begin{figure}[htbp]
\centerline{\includegraphics[width=0.5\textwidth]{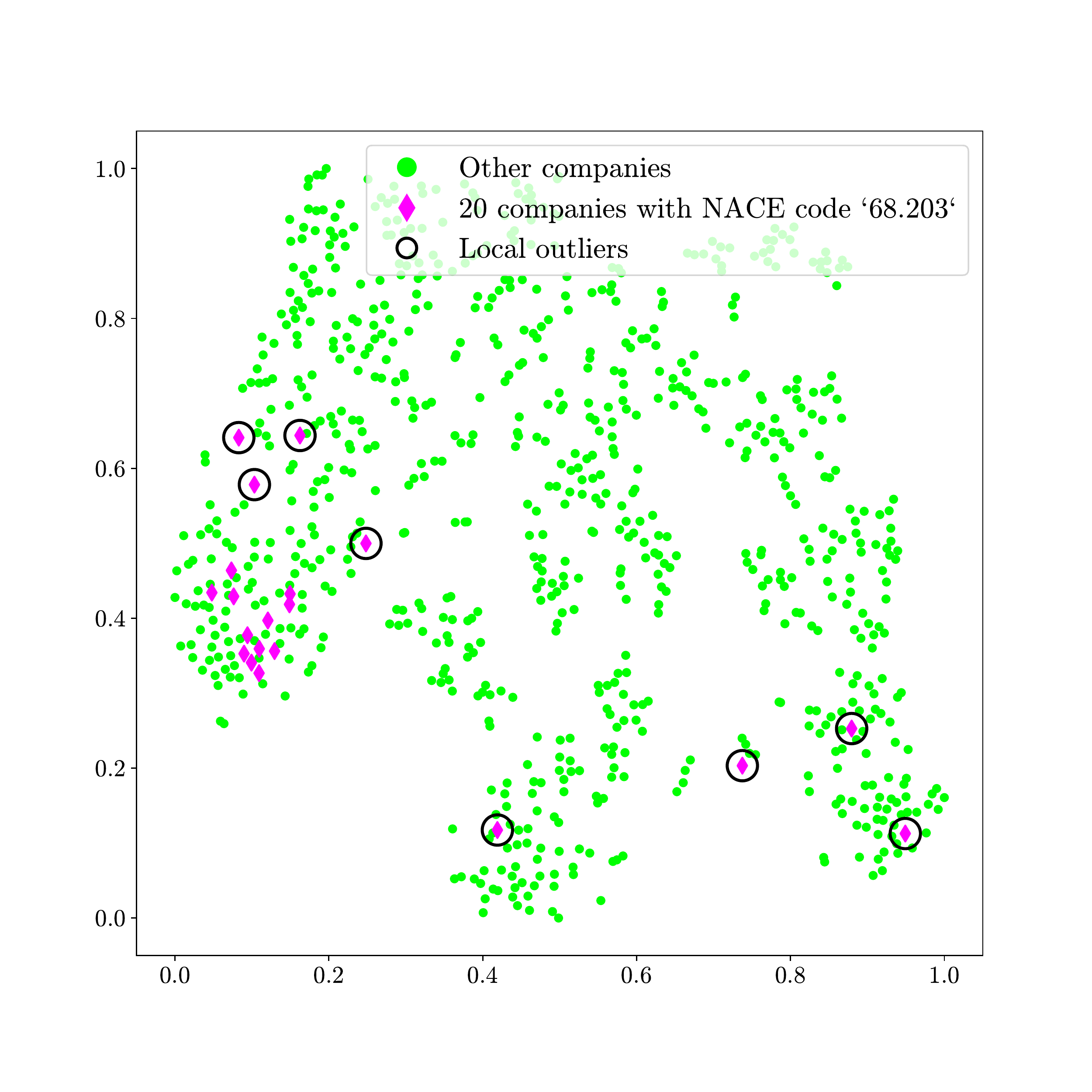}}
\caption{Two-dimensional company visualisation with t-SNE where the inter-company distances are computed by the EMD-GDM instead of the Euclidean distance. Furthermore, 20 companies involved in the rental and operation of non-residential real estate are visualized.}
\label{fig-tsne-industry}
\end{figure}

The second part of this experiment focuses on visualizing a specific industry in this two-dimensional company space. The visualized NACE code is ‘$68.203$’, which stands for \emph{rental and operation of own or leased non-residential real estate}. Figure \ref{fig-tsne-industry} shows this two-dimensional company representation. The pink diamonds represent the companies with industry code ‘$68.203$’, the other companies are represented by green dots. The black circles represent local industry outliers, detected by the LOF-algorithm. This anomaly detection method only considers the pairwise distances between the companies within this specific industry. In figure \ref{fig-tsne-industry} we can clearly see a group of companies with industry code ‘$68.203$’ that are grouped together. All the other companies within this industry are considered outliers based on the LOF-algorithm. We assume that the group of closely located companies within the same industry have similar balance sheets. The other companies within this industry should have dissimilar balance sheets. 

Subsequently, we want to qualitatively asses if the group of companies that are located closely together have similar balance sheets. We do this by inspecting their balance sheets and comparing them with the industry outliers. First, we consider the similar companies, where we compare the 5 companies with the lowest LOF-scores. 

Based on this qualitative evaluation we can confirm that the most similar companies have very similar balance sheet structures and distributions. Because a balance sheet can be divided into four separate trees (see section \ref{section4}), we discuss them separately. The debit active trees are highly similar between these companies. The largest weights are situated on the ledger accounts $221000$ and $222000$, which represent $buildings$ and $terrains$, respectively. Because these companies rent out real estate, this is also significantly tied to their industry activity. Since the highest ledger account weights are located on $buildings$ and $terrains$, the credit active trees are also comparable amongst these companies. The building- and terrain-related depreciation weights are the most significant weights of the credit active trees. The distances between the credit passive trees are all equal to zero, indicating that these ledger accounts have no weights. The debit passive trees tend to be the most different between these companies. This also makes sense; organizations that rent out property should have some property of their own, resulting in active trees that are fairly similar. This does not necessarily imply that their businesses are funded in the same way. Within a group of comparable businesses, their way of reserving profit differs as well as the amount of debt. One advantage is that the different types of profit reservation are located close to each other in the tree. This means that the distance between two profit-reserving companies is likely to be shorter than the distance between two companies that do not reserve profit. Nonetheless, a company's nearest neighbor tends to have comparable financial structures.

When we compare the industry outliers to the set of similar companies, we can see that there are significant differences. The active debit section no longer consists solely of $buildings$ and $terrains$; these companies also contain a lot of $installations$ and $financial\ assets$, resulting in a completely different amortization structure. Aside from that, their debit and credit passive trees show distinct variances. In comparison to the similar companies, we can see that the credit passive trees of the outliers have more distinct characteristics. The debit passive trees do not appear to be similar, as seen by their larger pairwise distances when compared to the group of similar companies.

We can confirm that our method further differentiates companies based on their financial statement information by taking into account structure and value information. This confirms the assumption that companies with a similar balance sheet structure can also be different. Companies within the same industry could have a very similar active structure, but this does not necessarily mean that their passive structure is. 

\section{Conclusion}\label{section6}

In this paper we propose a new graph distance metric that allows one to quantify the similarity between two financial statements. Unlike previous research, our method uses both structure and value information. The experimental results indicate that our proposed strategy outperforms the state-of-the-art methods for this task. This implies that integrating both structure and value information improves the utility of the distance metric. We also show that our method is able to find similar companies as well as company outliers based on their financial statement information.

In the future, this work could be extended in several directions. Since the optimized flows are traceable we could add an explainability layer that explains how companies are different. Another interesting direction is to embrace the dynamic nature of financial statement by representing companies as a time series of vertex-weighted trees. This means that the distance metric would also be dependant on the history of a company. Finally, we like to see the usefulness of our proposed metric in other industries, where data can be modeled as a vertex-weighted trees such as bioinformatics or social media. 

\section*{Acknowledgment}
This research received funding from the Flemish Government, through Flanders Innovation \& Entrepreneurship (VLAIO, project HBC.2020.2883) and from the Flemish Government under the “Onderzoeksprogramma Artificiële Intelligentie (AI) Vlaanderen” programme. We also wish to acknowledge the feedback provided by Nick Meerlaen, an accounting specialist of Silverfin.

\bibliographystyle{IEEEtran}
\bibliography{mvc.bbl}

\end{document}